\documentclass[aip,jap,reprint,twocolumn]{revtex4-1}
\usepackage{graphicx}
\usepackage{amsmath, amsthm, amssymb}
\usepackage[usenames,dvipsnames]{xcolor}

\def\l{\left}
\def\r{\right}
\def\benum#1\eenum{\begin{enumerate}#1\end{enumerate}}
\def\be#1\ee{\begin{equation}#1\end{equation}}
\def\bml#1\eml{\begin{multline}#1\end{multline}}
\def\ba#1\ea{\begin{align}#1\end{align}}
\def\bas#1\eas{\begin{align*}#1\end{align*}}
\def\eq#1{(\ref{eq:#1})}

\def\bit{\begin{itemize}}
\def\eit{\end{itemize}}
\def\l{\left}
\def\r{\right}

\def\bg#1\eg{\begin{gather}#1\end{gather}}

\newcommand{\B}[1]{\mathbf{#1}}

\def\bm#1\em{\begin{pmatrix}#1\end{pmatrix}}
\def\bg#1\eg{\begin{gather}#1\end{gather}}
\def\t{\text}
\def\n{\nonumber}

\begin{document}

\title{An analysis method for transmission measurements of superconducting resonators with applications to quantum-regime dielectric-loss measurements}

\author{Chunqing Deng}
\email{cdeng@uwaterloo.ca}
\author{Martin Otto}
\author{Adrian Lupascu}
\affiliation{Institute for Quantum Computing, Department of Physics and Astronomy, and Waterloo Institute for Nanotechnology, University of Waterloo, Waterloo, Ontario N2L 3G1, Canada}

\date{\today}

\begin{abstract}
Superconducting resonators provide a convenient way to measure loss tangents of various dielectrics at low temperature. For the purpose of examining the microscopic loss mechanisms in dielectrics, precise measurements of the internal quality factor at different values of energy stored in the resonators are required. Here, we present a consistent method to analyze a LC superconducting resonator coupled to a transmission line. We first derive an approximate expression for the transmission S-parameter $S_{21}(\omega)$, with $\omega$ the excitation frequency, based on a complete circuit model. In the weak coupling limit, we show that the internal quality factor is reliably determined by fitting the approximate form of $S_{21}(\omega)$. Since the voltage $V$ of the capacitor of the LC circuit is required to determine the energy stored in the resonator, we next calculate the relation between $V$ and the forward propagating wave voltage $V_{\t{in}}^+$, with the latter being the parameter controlled in experiments. Due to the dependence of the quality factor on voltage, $V$ is not simply proportional to $V_{\t{in}}^+$. We find a self-consistent way to determine the relation between $V$ and $V_{\t{in}}^+$, which employs only the fitting parameters for $S_{21}(\omega)$ and a linear scaling factor. We then examine the resonator transmission in the cases of port reflection and impedance mismatch. We find that resonator transmission asymmetry is primarily due to the reflection from discontinuity in transmission lines. We show that our analysis method to extract the internal quality factor is robust in the non-ideal cases above. Finally, we show that the analysis method used for the LC resonator can be generalize to arbitrary weakly coupled lumped and distributed resonators. The generalization uses a systematic approximation on the response function based on the pole and zero which are closest to the resonance frequency. This Closest Pole and Zero Method (CPZM) is a valuable tool for analyzing physical measurements of high-Q resonators.
\end{abstract}

\maketitle

\section{Introduction}
Measuring the response of a resonator is a generic method for measurements of properties of solids at low temperature, such as surface resistance\cite{oates1990measurement} of superconductors and complex permittivity\cite{Krupka1999} of dielectrics. It is also very useful for sensitive detection of photons\cite{Day2003}, states of superconducting quantum bits\cite{Vijay2011}, and the zero-point motion of mechanical oscillators\cite{Regal2008}. Recently, the need for increasing coherence times of superconducting qubits motivated new interest in searching for low-loss dielectric materials as well as understanding their loss mechanism. Internal quality factors of several types of superconducting thin-film resonators\cite{Paik2010, Geerlings2012, Megrant2012, OConnell2008, Weber2011, Cicak2010, Vissers2010, Macha2010, Gao2007, Barends2010} have been measured in detail. The most common way to measure the internal quality factor is to measure the S-parameters of resonant circuits. For this purpose, a number of methods were developed to extract the quality factor of a single resonator from reflection or transmission measurements\cite{Zmuidzinas2012}. In order to measure multiple resonators simultaneously, methods\cite{Wisbey2010, Khalil2012, Megrant2012} based on frequency multiplexing have been developed. However, all the previous methods derive the scattering parameters (S-parameters)\cite{Pozar2009} by using equivalent coupled circuits in the vicinity of the resonance. In this paper, we present a consistent method to approximate the exact S-parameters by analyzing the algebraic properties of the S-parameter function. We start from a circuit model in which a LC resonator couples to a transmission line both inductively and capacitively. We develop the Closest Pole and Zero Method (CPZM) to reduce the transmission S-parameter $S_{21}(\omega)$, with $\omega$ the excitation frequency, of the circuit to a ratio of complex linear functions. The approximated $S_{21}(\omega)$ only depends on four parameters: resonance frequency $\omega_0$, internal quality factor $Q_i$, external quality factor $Q_e$ and a parameter $Q_\alpha$ which is related to the resonance asymmetry. The internal quality factor $Q_i$ can be extracted from fitting the data to the simplified $S_{21}(\omega)$ function as opposed to the full circuit-dependent form.

We also discuss another important aspect of dielectric loss characterization. In order to understand sources and mechanisms of the loss, the dependence of dielectric loss on energy stored in the resonator is required. The energy dependence of loss can be characterized\cite{Paik2010, Weber2011, Martinis2005} by measuring $Q_i$ as a function of the voltage $V$ across the capacitor in the LC resonator. However, this is a nontrivial task since one usually does not have precise knowledge of all the circuit parameters needed to calculate $V$. We show that the ratio $V/V_{\t{in}}^+$, where $V_{\t{in}}^+$ is the forward propagating wave amplitude at the excitation port, can be calculated using parameters determined from the fit of the $S_{21}(\omega)$ function and an additional scaling factor. The scaling factor depends on circuit parameters which cannot be directly determined from the fit. However, it can be estimated based on circuit parameter simulations. An important point to note is that the knowledge of the voltage $V$ up to a fixed factor is still valuable for comparing experimental results with theoretical predictions of the loss mechanism.

Although the reduced $S_{21}$ fits the experimental data extremely well, in some measurements the resonance curve shows a pronounced asymmetry, which is inconsistent with realistic values of circuit parameters. We analyze a more complete model which takes the reflection of wire bonds and impedance mismatch into account. We find that for realistic circuits the asymmetric resonance primarily arises from wave reflection by the wire bonds used to connect the device and the measurement apparatus. We show that despite these non-idealities, our procedure for extracting $Q_i$ based on the simplified $S_{21}$ form is still valid.

We note that, in previous work, simplified equivalent circuits were used to obtain a simple expression for the S-parameters in the linear fractional form\cite{Kajfez1994}. However, in this paper, we develop a systematic approach to reduce the response function of an arbitrary resonator to a linear fractional form by using specifically its algebraic properties.

This paper is organized as follows. In Section~\ref{sec:S21}, we approximate the transmission S-parameter $S_{21}$ of the resonance circuit we consider to a linear fractional form by applying the CPZM. In Section~\ref{sec:VVin}, we investigate the complex function $v(\omega) = V/V_{\t{in}}^+$, where $V$ is the capacitor voltage and $V_{\t{in}}^+$ is the forward propagating wave voltage. We approximate $v(\omega)$ to a linear fractional form using the same method and we find that $v$ depends on $Q_i$ in general. We derive a formula to transform the points $(V_{\t{in}}^+, Q_i)$ to $(V, Q_i)$. In Section~\ref{sec:asymmetry}, we employ the CPZM to analyze the transmission parameter $S_{21}$ of a circuit where we consider the inductance of the bonding wires and the effect of variation of impedance of different transmission lines. In Section~\ref{sec:generalize}, we discuss the general application of CPZM to the circuit response of both lumped and distributed high-Q resonators.

\section{Derivation of resonator transmission} \label{sec:S21}

\subsection{Experimental design and circuit model}
\begin{figure}[b]
\includegraphics[width=8.5cm]{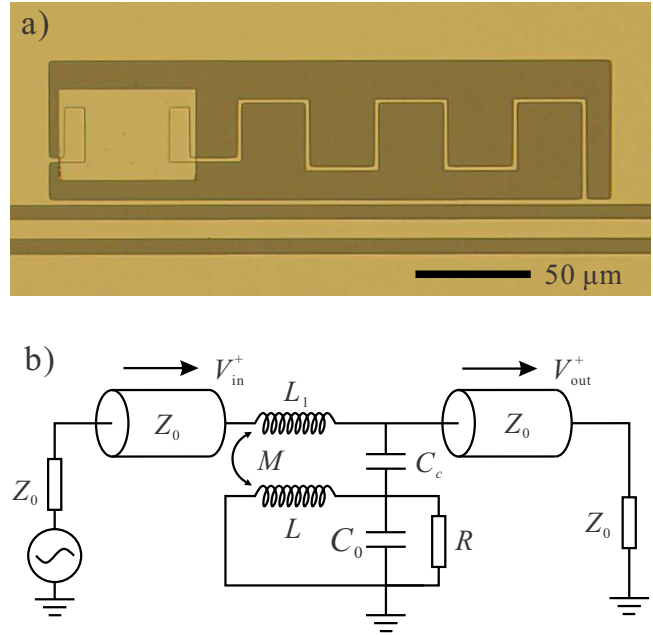}
\caption{(a) Optical microscope image of an LC resonator coupled to a transmission line, designed for measuring the loss tangent of the dielectric in the parallel-plate capacitor. The device is fabricated using two aluminum layers, patterned with electron-beam lithography. The dielectric is a 5~nm thick layer of AlO$_x$. The oxide is grown by applying oxygen plasma onto the bottom aluminum layer. The nominal capacitance value is 2.5~pF. (b) Lumped element circuit model for the device in (a).}
\label{fig:sample}
\end{figure}

We consider a LC resonator coupled to a transmission line. In our experiments, the resonator consists of a meander inductor $L$ and a parallel capacitor $C_0$. The transmission line is a coplanar waveguide (CPW) (see Fig.~\ref{fig:sample}(a) for the sample design). We use a model circuit similar to that proposed in Ref.~\onlinecite{Khalil2012}, as shown in Fig.~\ref{fig:sample}(b). The LC resonator couples to the transmission line through a mutual inductance $M$ and a capacitor $C_c$. Dielectric loss is modeled by a resistor $R$ shunting the lumped capacitor $C_0$. Both the input and output transmission lines have characteristic impedance $Z_0$.\\

\subsection{Exact $S_{21}$ from circuit model}
\begin{figure}[!]
\includegraphics[width=8.5cm]{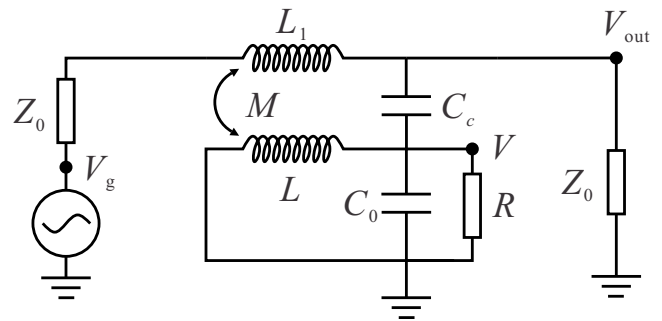}
\caption{Equivalent circuit of the circuit showed in Fig.~\ref{fig:sample}(b).}
\label{fig:eqcircuit}
\end{figure}

Fig.~\ref{fig:eqcircuit} shows the equivalent circuit of the network shown in Fig.~\ref{fig:sample}(b). Since the impedance of the source and load match the characteristic impedance of the transmission line, we have $V_g = 2V_{\t{in}}^+$ and $V_{\t{out}}^+ = V_{\t{out}}$ where $V_g$ is the voltage of the source and $V_{\t{out}}$ is the voltage on the output load. We use Kirchhoff's laws to calculate $S_{21} = V_{\t{out}}^+/V_{\t{in}}^+ = 2V_{\t{out}}/V_g$, which gives:
\be
S_{21} = \frac{1}{1+\frac{Z_{\t{in}}'}{Z_{0}}+j\omega C_c Z_{\t{in}}'}\l( 2 + \frac{V}{V_{\t{in}}^+}\l(j\omega C_c Z_{\t{in}}' + \frac{M}{L}\r) \r) \label{eq:S21}
\ee
where
\begin{widetext}
\be
\frac{V}{V_{\t{in}}^+} = -2\left. \frac{\frac{M}{L}-j\omega C_c Z_{\t{out}}'}{Z_{\t{out}}'+Z_{\t{in}}'} \right/ \left( \frac{1}{j\omega L}+ \frac{1}{Z_C} + \frac{j\omega C_c}{j\omega C_c Z_{0} + 1} + \frac{(\frac{M}{L}-j\omega C_c Z_{\t{out}}')^2}{Z_{\t{out}}' + Z_{\t{in}}'} \r) \label{eq:VVin}
\ee
\end{widetext}
and
\bg
Z_{\t{in}}' = Z_{0} + j\omega L_1 - j\omega M^2/L , \\
Z_{\t{out}}' = \frac{Z_{0}}{1 + j \omega C_c Z_{0}}, \\
\t{and} \quad Z_{C} = \frac{1}{j\omega C_0 + 1/R}.
\eg

\subsection{Reduction of $S_{21}$ to a linear fractional form}
The S-parameter $S_{21}$ in Eq.~\eq{S21} is in a polynomial fractional form of $\omega$. Our goal is to derive, in a systematic manner, a simpler approximate form of the $S_{21}$. Since $S_{21}$ is dimensionless, it is convenient to express it in terms of a set of dimensionless parameters. We choose the following dimensionless parameters:
\be
\begin{split}
\alpha = \frac{M}{L},\quad \beta = \frac{L_1}{L}, \quad \gamma = \frac{C_c}{C_0}, \quad \xi = \frac{\sqrt{L/C_0}}{Z_0}, \\q = 1/Q = \frac{1}{R \omega_0' C_0}, \quad
\text{and}\quad x = 2R \omega_0' C_0\l(\frac{\omega}{\omega_0'}-1\r) \label{eq:dimlessPara}.
\end{split}
\ee
Here $\omega_0' = 1/\sqrt{L C_0}$ is the resonance frequency of the uncoupled resonator. Coupling to the transmission line will result in a shift in the resonance frequency from $\omega_0'$. When expressed in terms of the new parameters, the expression for $S_{21}$ becomes:
\be
S_{21} = \frac{f(x)}{g(x)} = \frac{a_0 + a_1 q x+ a_2 (q x)^2}{b_0 + b_1 q x+ b_2 (q x)^2 + b_3 (q x)^3 + b_4 (q x)^4} \label{eq:S21dimless}
\ee
where $f(x)$ and $g(x)$ are polynomial functions of $qx$. The expressions for the coefficients in Eq.~\eq{S21dimless} are given in Appendix~\ref{sec:appendix}.

The dimensionless parameters in Eq.~\eq{dimlessPara} are chosen such that their values are much smaller (i.e. $\alpha$, $\beta$, $\gamma$, and $q$) or comparable (i.e. $x$ and $\xi$) to unity, which allows for developing systematic order expansions. To illustrate this, in Table~\ref{tab:realpara}, we list realistic values of circuit parameters for a resonator realized in our experiments with an estimated bare resonance frequency $\omega_0' = 2\pi \times 5.9$~GHz. Corresponding values of the reduced parameters are shown in Table~\ref{tab:reducepara}.
\begin{table}[h]
\caption{Parameters of the circuit shown in Fig.~\ref{fig:eqcircuit}.\label{tab:realpara}}
\begin{tabular}{c c c c c c c }
  \hline \hline
  $L_1$ & $M$ & $L$ & $C_c$ & $C_0$ & $Z_0$ & $R$ \\ \hline
  0.71~pH & 11.9~pH & 288.7~pH & 5.0~fF & 2.5~pF & 50~$\Omega$ & 50~$k\Omega$\\
  \hline \hline
\end{tabular}
\caption{Reduced parameters.\label{tab:reducepara}}
\begin{tabular}{c c c c c c c }
  \hline \hline
  $\alpha$ & $\beta$ & $\gamma$ & $\xi$ & $q$   \\ \hline
  0.041 & 0.0025 & 0.0020 & 0.21 & 0.00021 \\
  \hline \hline
\end{tabular}
\end{table}
We note that while all the reduced parameters are small, $q$ may be smaller than one by many orders of magnitude (the quality factor of a superconducting resonator often reaches $10^6$ especially when driven at high power). For resonators weakly coupled to the transmission line, we expect that the resonance frequency and the width of the resonance are not perturbed very strongly. For this reason, the relevant range of the variable $x = 2Q \frac{\omega - \omega_0'}{\omega_0'}$ corresponds to $|x| \lesssim 1$.

\begin{figure}[]
\includegraphics[width=8.5cm]{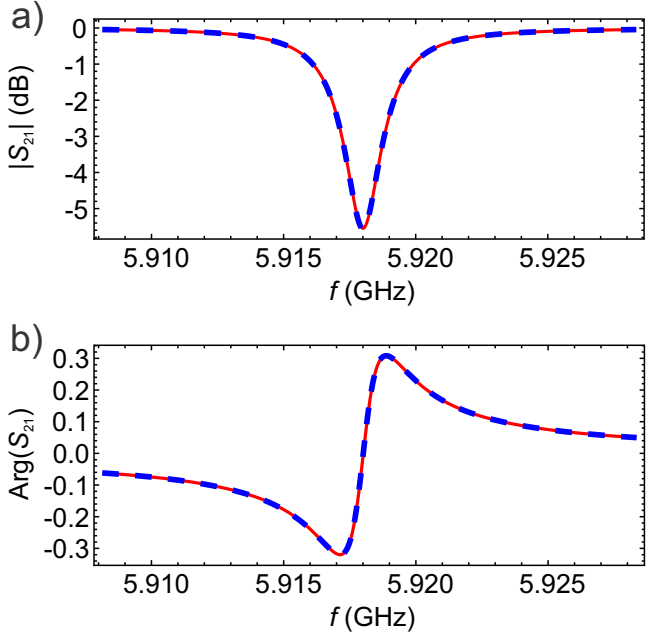}
\caption{Amplitude (a) and phase (b) of the exact $S_{21}$ (red solid line) and the reduced linear fractional form $\tilde S_{21}$ (blue dashed line). The expressions are defined in Eqs.~\eq{S21dimless} and \eq{S21approx}, respectively.}
\label{fig:approxS21}
\end{figure}

Since $q x$ is a number much smaller than the rest of the parameters, we can approximate $f(x)$ and $g(x)$ by keeping only the terms up to the first order of $q x$, which gives:
\be \label{eq:S21approx}
\tilde{S}_{21} = \frac{\tilde{f}(x)}{\tilde{g}(x)} = \frac{a_0 + a_1 qx}{b_0 + b_1 qx}.
\ee
Fig.~\ref{fig:approxS21} illustrates the accuracy of the approximation above by comparing both the amplitude and phase of $S_{21}$ and $\tilde{S}_{21}$ using realistic circuit parameters given in Table~\ref{tab:realpara}. Noting that both $\tilde{f}(x)$ and $\tilde{g}(x)$ contain terms up to the first order in $x$, we rearrange the terms and write $\tilde{S}_{21}$ in the following linear fractional form:
\be
\tilde{S}_{21} = |A_0|e^{j\phi}\frac{a + j x}{b + j x}. \label{eq:S21lf}
\ee
where $a = \frac{j}{q} \frac{a_0}{a_1}$, $b = \frac{j}{q} \frac{b_0}{b_1}$, $|A_0|e^{j\phi} = a_1/b_1$. We note that the parameter $|A_0|e^{j\phi}$ has a value which is very close to 1. For this reason it would be difficult to distinguish this factor from attenuation and gain factors in the measurement setup. Therefore, we simply drop this factor from now on. The expressions for $a$ and $b$ are ratios of polynomials, depending on all the reduced parameters, given in Appendix~\ref{sec:appendix}. We rewrite next the parameters $a$ and $b$ in a way which makes the dependence on $q$ explicit. We have:
\be
a = A + j\frac{B}{q},
\ee
and
\be
b = C + \frac{D}{q} + j(E + \frac{F}{q}),
\ee
where $A$, $B$, $C$, $D$, $E$ and $F$ depend on the reduced parameters other than $q$.

We simplify the coefficients $A$, $C$, and $E$ by using a Taylor expansion in terms of $\alpha$ and $\gamma$, which characterizes the resonator coupling and are therefore very small. We obtain:
\ba
A =& 1 + O(\alpha, \gamma), \\
C =& 1 + O(\alpha, \gamma), \\
E =& O(\alpha, \gamma).
\ea
For the purpose of order comparisons, we also write down the series expansion of $B$, $D$, and $F$ below:
\ba
B &= \gamma -\alpha\gamma-\gamma^2+ \cdots, \\
D &= \frac{(2+\beta  \xi ^2+\beta ^2 \xi ^2)}{\xi(4+\beta ^2 \xi ^2)}\gamma ^2 +\frac{2 \xi }{4+\beta ^2 \xi ^2}(\alpha^2-\beta\alpha\gamma)+\cdots, \\
F &= \gamma + \frac{-8+2 \beta -3 \beta ^2 \xi ^2}{2(4+\beta ^2 \xi ^2)}\gamma ^2 -\frac{4 \gamma \alpha+\beta\xi^2\alpha^2}{4+\beta ^2 \xi ^2} + \cdots .
\ea
These expansions are carried out using \emph{Mathematica}\cite{mathematica}. We find that $B$, $D$, and $F$ are at least in the second order of $\alpha$ and $\gamma$. For high-Q resonators($q \ll \alpha, \gamma$), we only retain the zeroth order terms for $A$, $C$, $E$. First order terms for $A$, $C$, and $E$ are not needed, due to the large $B/q$, $D/q$, and $F/q$ terms. With these approximations, the function $\tilde S_{21}$ takes the form:
\be
\tilde{S}_{21,\t{red}} = \frac{1+j\frac{B}{q}+jx}{1+\frac{D}{q}+j\frac{F}{q}+jx}. \label{eq:S21red}
\ee

The resonance curve described by Eq.~\eq{S21red} is asymmetric in general. We identify the resonance frequency as the value when $|\tilde{S}_{21,\t{red}}|$ is minimized. The resonance frequency $\omega_0$ depends on the internal loss of the capacitor (details shown in Appendix~\ref{sec:appendix}). However, for high-Q resonators, this dependence is weak for $q \ll 0$. For simplicity, we take the resonance frequency as the resonance frequency without internal loss ($q=0$) as the following:
\be
\omega_0 = \omega_0' (1-B/2), \label{eq:omega0'}
\ee
where
\be
B = \frac{\gamma(1-\alpha)}{1+\gamma-\alpha\gamma}.\label{eq:B}
\ee
The internal quality factor is defined as
\ba
Q_i =& \omega_0 R C_0 = \omega_0' (1-B/2)R C_0  \label{eq:Qi}\\
=& \frac{1-B/2}{q}.
\ea
We rewrite Eq.~\eq{S21red}, showing specifically the resonance frequency $\omega_0$ and the internal quality factor $Q_i$ to
\be
\tilde{S}_{21,\t{red}} = \frac{1+2jQ_i\frac{\omega - \omega_0}{\omega_0}}{1+\frac{Q_i}{Q_e}+j\frac{Q_i}{Q_\alpha}+2jQ_i\frac{\omega - \omega_0}{\omega_0}}, \label{eq:S21fitting}
\ee
with $Q_\alpha$ and $Q_e$ defined as the following expressions:
\ba
Q_\alpha =& \frac{1-B/2}{F-B}, \label{eq:Qalpha}\\
Q_e =& \frac{1-B/2}{D}. \label{eq:Qe}
\ea
The reduced $S_{21}$ in Eq.~\eq{S21fitting} is in a form which is similar to a Lorentzian. $Q_\alpha$ is a parameter that characterizes the asymmetry of $S_{21}$ with respect to the resonance frequency. In the limit where $Q_{\alpha} \rightarrow \infty$, $\omega_0 (Q_i^{-1}+Q_e^{-1})$ is the width of the transmission signal in the frequency domain. Therefore, $Q_e$ is the external quality factor of the resonator.

Eq.~\eq{S21fitting} is the main result of this section. By fitting Expression~\eq{S21fitting} to $S_{21}$ data, we can uniquely determine the fitting parameters $\omega_0$, $Q_i$, $Q_e$, and $Q_\alpha$. It is also worth to point out that Eq.~\eq{S21fitting} has the same form as to Eq.~(3) in Ref.~\onlinecite{Megrant2012} and it can be transfer into Eq.~(13) in Ref.~\onlinecite{Khalil2012} by redefining parameters as the following:
\bg
\widetilde{\omega}_{0} = \omega_{0}\left(1-\frac{1}{2Q_{\alpha}}\right), \n\\
\widetilde{Q}_{i} = Q_i \left(1-\frac{1}{2Q_{\alpha}}\right),\n\\
\widetilde{\hat{Q}}_{e}^{-1} = 1/Q_e + j/Q_\alpha, \n\\
\widetilde{Q}^{-1} = 1/Q_e + 1/Q_i,
\eg
where $\widetilde{\omega}_{0}$, $\widetilde{Q}_{i}$, $\widetilde{\hat{Q}}_{e}$, and $\widetilde{Q}$ are parameters defined in Ref.~\onlinecite{Khalil2012}. The difference in expressions of the resonance frequency is due to the fact that Ref.~\onlinecite{Khalil2012} considers the resonance frequency as the frequency at maximum $|V|$ while we consider it as the frequency at minimum $|S_{21}|$. In Section~\ref{sec:generalize}, we discuss the more general arguments that justify our systematic approach. We show there that this method approximates the S-parameter from a complex function with multiple poles and zeros to a reduced form governed by only one pole and one zero whose real parts are the closest to the resonance frequency. We call this method the Closest Pole and Zero Method (CPZM).

\section{Voltage on lumped element capacitor} \label{sec:VVin}
Characterizing the field dependence of dielectric loss is essential in understanding the loss mechanism of dielectrics. For amorphous dielectric materials, tunneling two-level systems (TTLS)\cite{Phillips1999, Martinis2005} are believed to constitute the dominant source of loss in the quantum regime. The TTLS model has detailed predictions of the loss dependence on the AC electric field in the dielectric. In a LC superconducting resonator, the electric field in the dielectric is proportional to the voltage $V$ on the capacitor.

Using circuit theory, we can determine $V(Q_i,V_{\t{in}}^+)$, with $Q_i$ the resonator internal quality factor and $V_{\t{in}}^+$ the voltage of the forward propagating wave towards the resonator input port. If $Q_i$ is a constant or the circuit is strongly coupled to the transmission line, $V$ is proportional to $V_{\t{in}}^+$. The experimentally measured $(Q_i,V_{\t{in}}^+)$ dependence is transformed into a $(Q_i,V)$ dependence assuming a constant ratio $V/V_{\t{in}}^+$. In this section we develop a systematic method to express the dependence of $V$ on $V_{\t{in}}^+$ in a way which takes into account the voltage dependence of the quality factor. We find that the $(Q_i,V)$ dependence determined in this way is different than the case when $V/V_{\t{in}}^+$ is simply assumed constant. The difference is more than a trivial scaling factor, and has an effect on the exponents appearing in the voltage dependence of the loss for realistic circuits.

\begin{figure}[]
\includegraphics[width=8.5cm]{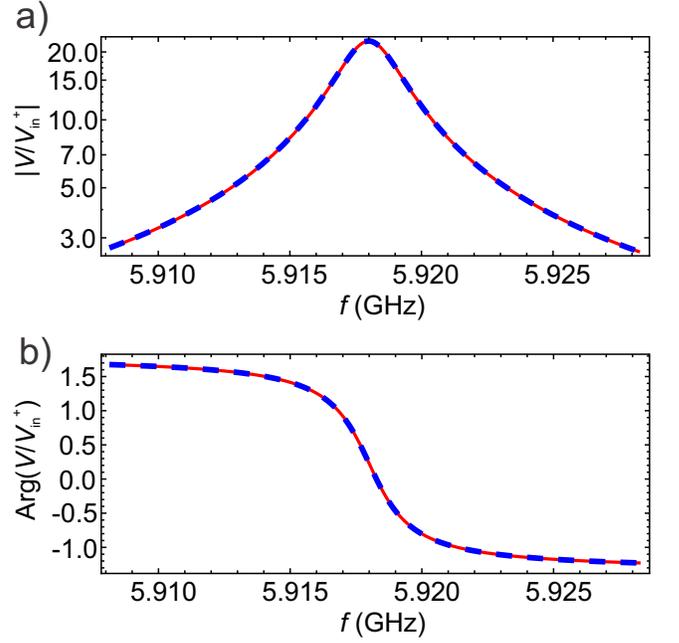}
\caption{Amplitude (a) and phase (b) of the exact $V/V_{\t{in}}^+$ (red solid line) and the reduced linear fractional form $\tilde v$ (blue dotted line).}
\label{fig:approxVVin}
\end{figure}

Starting from Eq.~\eq{VVin}, we introduce $v(\omega) = V/V_{\t{in}}^+$ and express it using the dimensionless parameters introduced in Eq.~\eq{dimlessPara}. In terms of the reduced parameters introduced earlier, $v$ is a high order polynomial fractional form $v = f_v(x)/g_v(x)$ (see Eq.~\eq{VVinred} in Appendix~\ref{sec:appendix} for the full expression of $v$). We apply our CPZM to approximate $v$ to $\tilde v$, a linear fractional form of $q$:
\be
\tilde{v} = \frac{\tilde f_v(x)}{\tilde g_v(x)} = \frac{c_0 + c_1 qx}{d_0 + d_1 qx}, \label{eq:VVindimless}
\ee
with $\tilde f_v(x)$ and $\tilde g_v(x)$ containing terms up to first order in $x$. Fig.\ref{fig:approxVVin} illustrates the accuracy of the approximation above by comparing both the amplitude and phase of $v$ and $\tilde v$ using realistic circuit parameters given in Table~\ref{tab:realpara}. We rewrite $\tilde v$ as
\be
\tilde v = \zeta \frac{a' + jx}{b' + jx},
\ee
where $\zeta = c_1/d_1$, $a' = \frac{j}{q}\frac{c_0}{c_1}$ and $b' = \frac{j}{q}\frac{d_0}{d_1}$. Separating $q$ from the rest of the reduced parameters gives
\be
\tilde v = \zeta \frac{K/q + jJ/q + jx}{H + N/q + j(O+P/q)+jx}.
\ee
Interestingly, some of the coefficients in the above equation are exactly the same as those in the $\tilde S_{21}$ formula given by Eq.~\eq{S21fitting}. We have $H = C$, $N = D$, $O = E$, and $P = F$. Since it is the features of the near-resonance transmission that allow one to extract $Q_i$, we calculate the value of $\tilde v(\omega)$ for $\omega = \omega_0$. At this frequency, $x = \frac{2}{q}(\omega_0/\omega_0'-1) = -\frac{B}{q}$. We obtain:
\be
\l. \frac{V}{V_{\t{in}}^+} \r\vert_{\omega = \omega_0} \approx \tilde v_0 = \frac{\lambda}{q_i + q_e + j q_\alpha} \label{eq:VVinapprox}
\ee
where $q_i = 1/Q_i$, $q_e = 1/Q_e$, $q_\alpha = 1/Q_\alpha$ and
\be
\lambda = \zeta\frac{K+j(J-B)}{1-B/2}. \label{eq:lambda}
\ee
$\lambda$ is a scaling factor which does not depend on internal quality factor of the resonator. The full form of $\lambda$ is given by Eq.~\eq{lambda_full} in Appendix~\ref{sec:appendix}.

To estimate this scaling factor, we write $\lambda$ as a Taylor series of small numbers $\gamma$, $\alpha$, and $\beta$ to their first orders, which gives
\be
\lambda = -\alpha\xi + j \gamma + \cdots,
\ee
where $\xi$ has a quantity which can be comparable to unity. We approximate the expression above further to $\lambda \approx -M\omega_0'/Z_0 + jC_c/C_0 \approx -M\omega_0/Z_0 + jC_c/C_0$. Given that $\omega_0$ can be obtained from fitting the $S_{21}$ data and $Z_0$ is usually assumed to be $50~\Omega$, we only need to estimate circuit parameters, $M$, $C_c$, and $C_0$, for determining $\lambda$.

\begin{figure}[]
\includegraphics[width=8.5cm]{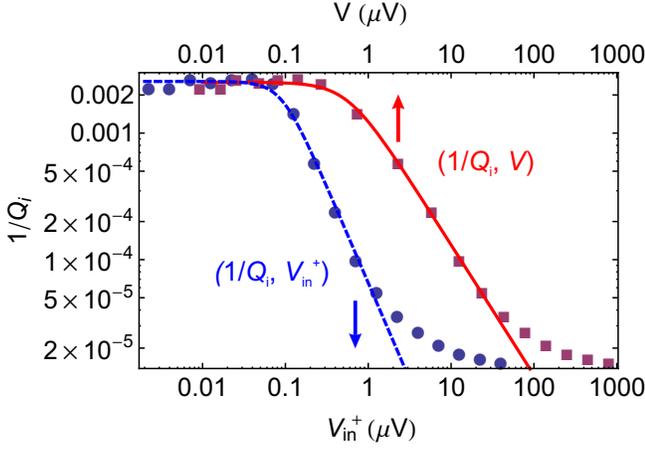}%
\caption{Experimentally measured loss dependence on $V_{\t{in}}^+$(blue dots) and on $V$(red squares). The red line is a result of fitting Eq.~\eq{TLS} to $(V, Q_i)$ which yields $\Delta = -0.0613$ and $Q_{i,0} = 397$. The blue dashed line is a result of fitting Eq.~\eq{TLS} to $(V_{\t{in}}^+, Q_i)$ which yields $\Delta = -1.05$ and $Q_{i,0} = 391$.}
\label{fig:qivsVin_exp}
\end{figure}

In experiments, one measures the $S_{21}$ at various values of $V_{\t{in}}^+$. For each value of $V_{\t{in}}^+$, a corresponding $Q_i$ is obtained from fitting Eq.~\eq{S21fitting} to the $S_{21}$ data. $Q_e$, $Q_\alpha$, and $\omega_0$ are obtained from the fit as well. Then one can use Eq.~\eq{VVinapprox} to transform data set $(V_{\t{in}}^+, Q_i)$ into $(V, Q_i)$. To illustrate the difference between the above two data sets, we plot in Fig.~\ref{fig:qivsVin_exp} the experimental $(V_{\t{in}}^+, Q_i)$ data of the parallel plate LC resonator shown in Fig.~\ref{fig:sample}(a) and the corresponding  $(V, Q_i)$ data transformed with Eq.~\eq{VVinapprox} with $Q_e = 1984$, $Q_\alpha = 4128$, $\omega_0 = 2\pi \times 7.665$ GHz, and $\lambda = 0.01146$. The dielectric layer between the paralleled plates is made of aluminum oxide, whose quantum regime loss is believed to be due to TTLSs. According to Ref.~\onlinecite{Martinis2005}, the TTLS loss at a constant temperature is predicted to be
\be
\frac{1}{Q_i} = \frac{1/Q_{i,0}}{\sqrt{1+(V/V_c)^{2 - \Delta}}} \label{eq:TLS}
\ee
with $\Delta = 0$. We fit Eq.~\eq{TLS} to both $(V, Q_i)$ and $(V_{\t{in}}^+, Q_i)$ data. We note that the fit to $(V, Q_i)$ yields $\Delta = -0.0613$, which is a very good agreement with the TTLS model while the fit to $(V_{\t{in}}^+, Q_i)$ yields $\Delta = -1.05$. As a result, the transformation from $V_{\t{in}}^+$ to $V$ with the loss of the resonator taken into account is crucial for determining the voltage dependence of dielectric loss. However, we find from Eq.~\eq{VVinapprox} that $V$ is approximately proportional to $V_{\t{in}}^+$ in the over-coupled limit, $Q_i \gg Q_e$. In this case, the fits to $(V, Q_i)$ and $(V_{\t{in}}^+, Q_i)$ will give approximately the same exponents $\Delta$.

\section{Explanation of asymmetric transmission} \label{sec:asymmetry}
\begin{figure}[b]
\includegraphics[width=8.5cm]{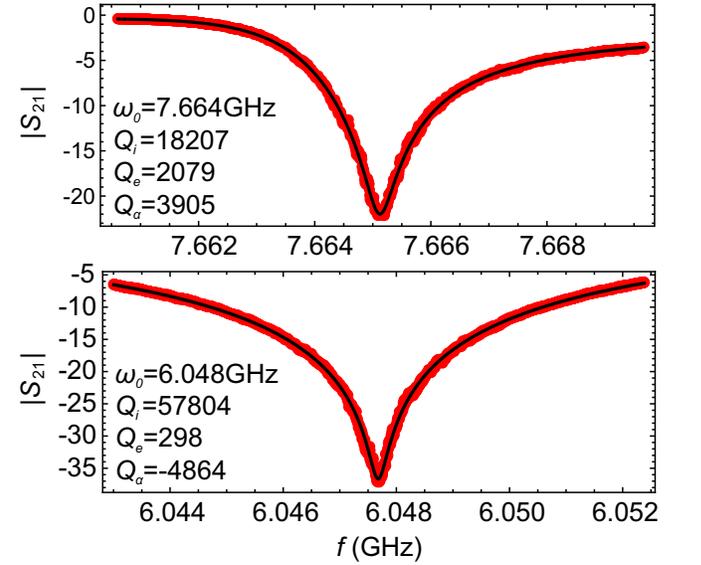}%
\caption{Measured transmission curves (red dots) and the fits using Eq.~\eq{S21fitting} (black line) for two resonators on the same chip showing different types of asymmetries. The extracted fit parameters are indicated in the graph legends.}
\label{fig:asymmetric_exp}
\end{figure}

\begin{figure*}[]
\includegraphics[width=17cm]{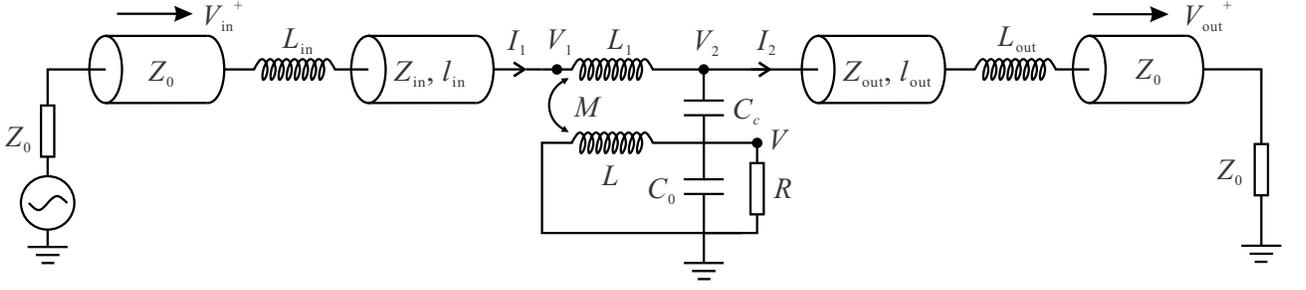}%
\caption{Circuit model of a LC resonator coupled to a non-ideal transmission line. The on-chip CPW transmission line close to the input/output port has length $l_{\t{in/out}}$ and characteristic impedance $Z_{\t{in/out}}$. They are connected to the input/output port through wire bonds model by inductance $L_{\t{in/out}}$. Both the input and output ports are terminated with an impedance $Z_0$.}
\label{fig:wirebondcircuit}
\end{figure*}

Although the function given in Eq.~\eq{S21fitting} fits the experimental data very well, the amount of transmission asymmetry characterized by $Q_\alpha$ cannot always be explained by the circuit model in Fig.~\ref{fig:sample}(b) with a set of realistic parameters. Fig.~\ref{fig:asymmetric_exp} shows two $|S_{21}|$ curves, of LC resonators measured in our experiments, which are highly asymmetric. A similar, large, asymmetry was observed in experiments of Geerlings et al.\cite{Geerlings2012} and Megrant et al\cite{Megrant2012}. Khalil et al.\cite{Khalil2012} attribute this effect to an on-chip inductor and transmission line impedance mismatches. Our experiments indicate a very large transmission line impedance mismatch is required to explain the asymmetry transmission, which is unlikely to occur in our setup. Motivated by this observation, we analyze a more complete circuit model, shown in Fig.~\ref{fig:wirebondcircuit}. In this model we take into account the connections between transmission lines on chip and outside the chip, implemented with bonding wires. We model these using inductances $L_{\t{in}}$ and $L_{\t{out}}$. These connections act as reflection points for the propagating microwaves. The two parts of the on-chip transmission lines, between the bonding connection and the resonator, are assumed to have characteristic impedance $Z_{\t{in/out}}$ and length $l_{\t{in/out}}$ respectively.

For such a circuit, consisting of several two-port components, it is convenient to use the ABCD transmission matrix method\cite{Pozar2009}. The ABCD matrix $\B{T}$ of a general 2-port circuit is defined as
\be
\left(
  \begin{array}{c}
    V_{\t{in}} \\
    I_{\t{in}} \\
  \end{array}
\right)
= \B{T}
\left(
  \begin{array}{c}
    V_{\t{out}} \\
    I_{\t{out}} \\
  \end{array}
\right)
\ee
where $V_{\t{in/out}}$ and $I_{\t{in/out}}$ are the voltages and currents at the input and output port respectively. We first use Kirchhoff's law to calculate the ABCD transmission matrix of the resonator part of the circuit (see Fig.~\ref{fig:wirebondcircuit} for the circuit schematics). We write down the circuit equations which relate the voltages, $V_1$ and $V_2$, and the currents, $I_1$ and $I_2$, to the capacitance voltage $V$ as follows:

\bg
V_{1} - V_{2} = j\omega L_1 I_{1} - j\omega M \frac{V + j\omega M I_{1}}{j\omega L}, \\
V_{2} - V = - \frac{1}{j \omega C_c}(I_{2}-I_{1}), \\
\frac{V(j\omega R C_0+1)}{R}+\frac{V+j\omega M I_{1}}{j\omega L} = I_{1} - I_{2}.
\eg
Based on these equations we can derive the ABCD matrix for the circuit
\be
\B{T}_{\t{res}} =
\left(
  \begin{array}{cc}
    T_A & T_B \\
    T_C & T_D \\
  \end{array}
\right)
\ee
where
\begin{widetext}
\ba
T_A =& \frac{-j \omega  \left(L-C_c L L_1 \omega ^2+C_c M^2 \omega ^2\right)+R \left(-1+(C_0 L+{C_c} (L+{L_1}-2 M)) \omega ^2+{C_0} {C_c} \left(-L {L_1}+M^2\right) \omega ^4\right)}{-j L \omega +R \left(-1+(({C_0}+{C_c}) L-{C_c} M) \omega ^2\right)}, \\
T_B =& \frac{\omega  \left(-j {L_1} R+\left(L {L_1}-M^2\right) \omega +j ({C_0}+{C_c}) \left(L {L_1}-M^2\right) R \omega ^2\right)}{-j L \omega +R \left(-1+(({C_0}+{C_c}) L-{C_c} M) \omega ^2\right)}, \\
T_C =& \frac{{C_c} \omega  \left(L \omega +j R \left(-1+{C_0} L \omega ^2\right)\right)}{-j L \omega +R \left(-1+(({C_0}+{C_c}) L-{C_c} M) \omega ^2\right)}, \\
T_D =& \frac{-j L \omega +R \left(-1+({C_0}+{C_c}) L \omega ^2\right)}{-j L \omega +R \left(-1+(({C_0}+{C_c}) L-{C_c} M) \omega ^2\right)}.
\ea
\end{widetext}
For the other parts of the circuit, the ABCD matrixes of the wire bond inductors are
\be
\B{T}_{L_{\alpha}} =
\left(
  \begin{array}{cc}
    1 & j\omega L_{\alpha} \\
    0 & 1 \\
  \end{array}
\right)
\ee
and the ABCD matrixes for the two pieces of the transmission lines between wire bonds and the resonator are
\be
\B{T}_{TL_{\alpha}} =
\left(
  \begin{array}{cc}
    \cos \beta_{\alpha} l_{\alpha} & j Z_{\alpha} \sin \beta_{\alpha} l_{\alpha} \\
    j \sin \beta_{\alpha} l_{\alpha} /Z_{\alpha} & \cos \beta_{\alpha} l_{\alpha} \\
  \end{array}
\right),
\ee
where $\alpha$ can stand for ``in'' or ``out'', $\beta_\alpha = \omega \sqrt{\tilde L_\alpha \tilde C_\alpha}$, and $\tilde L_\alpha$ and $\tilde C_\alpha$ are the characteristic inductance and capacitance of the transmission line. We define the time, for microwave to propagate $l_\alpha$ in distance, as $t_\alpha = l_\alpha\sqrt{\tilde L_\alpha \tilde C_\alpha}$. Given the width of the resonance $\delta \omega \ll 2\pi/t_\alpha$, we can approximate $\B{T}_{\t{TL}_\alpha}$ to an $\omega$-independent ABCD matrix
\be
\tilde {\B{T}}_{\t{TL}_\alpha} =
\left(
  \begin{array}{cc}
    \cos \omega_0' t_\alpha & jZ_\alpha\sin\omega_0' t_\alpha \\
    j\sin\omega_0' t_\alpha/Z_\alpha & \cos \omega_0' t_\alpha \\
  \end{array}
\right)
\ee
in the vicinity of $\omega = \omega_0'$. In the end, we obtain the ABCD matrix of the complete circuit by multiplying all the ABCD matrices:
\ba
\B{T}_{\t{all}} =& \B{T}_{L_{\t{in}}} \B{T}_{\t{TL}_{\t{in}}} \B{T} \B{T}_{\t{TL}_{\t{out}}} \B{T}_{L_{\t{out}}} \\
\approx & \B{T}_{L_{\t{in}}} \tilde{\B{T}}_{\t{TL}_{\t{in}}} \B{T} \tilde{\B{T}}_{\t{TL}_{\t{out}}} \B{T}_{L_{\t{out}}}.
\ea
The scattering matrix element $S_{21}$ can be calculated from the ABCD matrix using the following equation:
\be
S_{21}' = \frac{2}{{T}_{\t{all},A} + {T}_{\t{all},B}/Z_0 +{T}_{\t{all},C}Z_0 + {T}_{\t{all},D}}, \label{eq:S21all}
\ee
where ${T}_{\t{all},a}$ with $a = A$, $B$, $C$, or $D$ is each of the matrix elements of $\B{T}_{\t{all}}$.

\begin{figure}[]
\includegraphics[width=8.5cm]{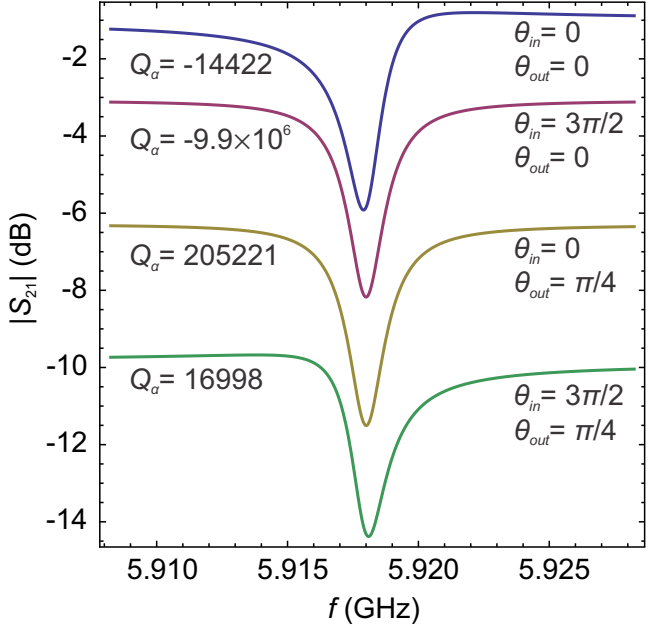}%
\caption{The asymmetric resonance calculated using different values of transmission line electrical length $t_{\t{in/out}}$. Here $\theta_{\t{in/out}} = \omega_0' t_{\t{in/out}}$. Each curve has an incremental offset of $-3$~dB. The asymmetry degree parameter $Q_\alpha$ is calculated for each curve. The symmetric $|S_{21}|$ case corresponds to $Q_\alpha \rightarrow \infty$.}
\label{fig:asymmetry}
\end{figure}

In Fig.~\ref{fig:asymmetry}, we plot $S_{21}'$ in Eq.~\eq{S21all} using $L_{\t{in}} = L_{\t{out}} = 370$~pH, $Z_{\t{in}} = Z_{\t{out}} = 50 \; \Omega$, and parameters in Table~\ref{tab:realpara}. These values of the inductance were determined based on numerical simulations of transmission through wire bonds with  a 1~mm length. Fig.~\ref{fig:asymmetry} shows that $S_{21}'$ with different $\theta_{\t{in}} = t_{\t{in}}\omega_0'$ and $\theta_{\t{out}} = t_{\t{out}}\omega_0'$ can have a different amount of asymmetry with different signs. In the multiplexed resonator measurement design, a number of resonators are coupled to the same transmission line. Since their positions relative to the reflection points (wire bonds here) and their resonance frequencies are different, the asymmetry feature of their transmission could also be different. This is what we observe in the experiment, which strengthens the hypothesis of our wire bond reflection model. The asymmetry is essentially due to reflections at the connection points which leads to weak standing waves with a length and frequency dependent pattern.

Finally, we apply our approximation procedure shown in Section~\ref{sec:S21} to this more complicated model and simplify the transmission $S_{21}'$ in Eq.~\eq{S21all}. We find that we reach a similar form to Eq.~\eq{S21red}, which is given by
\be
\tilde S_{21,\t{red}}' = \frac{1+j\frac{B}{q}+jx}{1+\frac{D'}{q}+j\frac{F'}{q}+jx}.
\ee
In the equation above, $B$ is exactly the same as in Eq.~\eq{B}. $D'$ and $F'$ are $q$-independent parameters. By using definition $\omega_0 = \omega_0' (1-B/2)$ and $Q_i = \omega_0 R C_0$ which are the same as Eq.~\eq{omega0'} and Eq.~\eq{Qi}, and definitions $Q_e' = (1-B/2)/(F'-B)$ and $Q_\alpha' = (1-B/2)/D'$ which are similar to Eq.~\eq{Qe} and \eq{Qalpha}, we obtain
\be
\tilde S_{21,\t{red}}' = \frac{1+2jQ_i\frac{\omega - \omega_0}{\omega_0}}{1+\frac{Q_i}{Q_e'}+j\frac{Q_i}{Q_\alpha'}+2jQ_i\frac{\omega - \omega_0}{\omega_0}} \label{eq:S21'fitting}
\ee
which has the same form as Eq.~\eq{S21fitting}. The parameters $Q_e'$ and $Q_\alpha'$ here are different from the corresponding $Q_e$ and $Q_\alpha$ in Eq.~\eq{S21fitting}. They in general depend on more circuit parameters, i.e. $L_{\t{in/out}}$, $Z_{\t{in/out}}$ and $l_{\t{in/out}}$, in addition to those circuit parameters in Table~\ref{tab:realpara}. By fitting Eq.~\eq{S21'fitting}, the intrinsic quality factor $Q_i$ and the resonance frequency $\omega_0$ can be determined reliably.

We emphasize that the same $Q_i$ and $\omega_0$ appear in both Eq.~\eq{S21fitting} and Eq.~\eq{S21'fitting} despite different external circuits coupled to the resonator. This shows that the impendence mismatch and wave reflection in the transmission line considered in this case does not compromise the determination of the internal quality factor $Q_i$.

\section{Generalization to other resonance circuits} \label{sec:generalize}
\begin{figure}[]
\includegraphics[width=8.5cm]{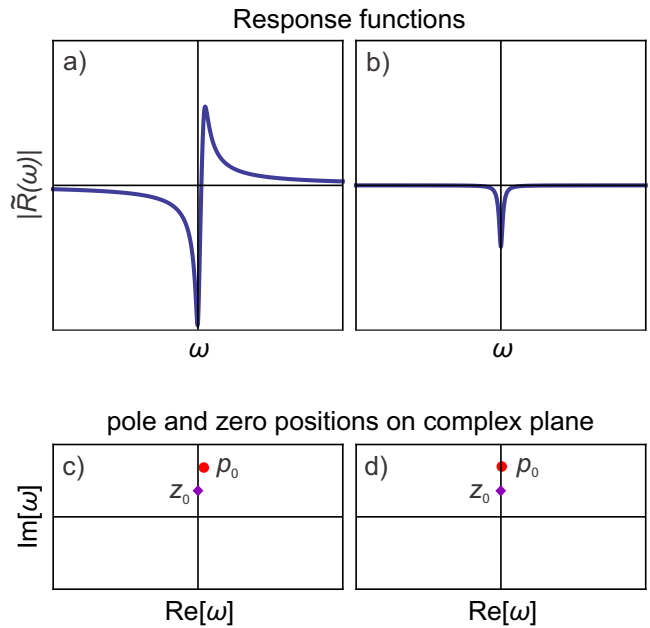}%
\caption{An asymmetric (a) and symmetric (b) response function with its pole and zero plotted in the complex plane in (c) and (d) respectively.}
\label{fig:pz}
\end{figure}

In this section we discuss the general application of CPZM to lumped and distributed resonators. We start with the consideration of lumped resonators. In this case, the response function (be it the transmission, reflection, or the ratio between voltages of any points) can be expressed as a polynomial ratio:
\be
R(\omega) = \frac{N(\omega)}{D(\omega)}, \label{eq:poly}
\ee
where $N(\omega)$ and $D(\omega)$ are polynomials, of degree $m$ and $n$ respectively. We can factor both polynomials to obtain
\be
R(\omega) = R_0 \times \frac{\prod_{i=0}^{n-1} (\omega-z_i)}{\prod_{j=0}^{m-1} (\omega-p_i)}
\ee
with $R_0$ a complex number and $z_i$ and $p_i$ zeros and poles of the response function $R(\omega)$. The entire resonance behavior of the circuit is described by the positions of the zeros and poles and the complex scaling factor. Let us consider the response of the resonator close to one of its bare resonances, at frequency $\omega_0'$. If the resonator is weakly coupled to its measurement setups, we expect that the resonant behavior will be reflected by a zero and/or a pole of $R(\omega)$ which are close to $\omega_0'$. We denote the closest zero and pole by $z_0$ and $p_0$. The response function can be expressed in the following form:
\be
R(\omega) = \frac{\bar{N}(\omega)(\omega - z_0)}{\bar{D}(\omega)(\omega - p_0)},
\ee
where $\bar{N}(\omega)$ and $\bar{D}(\omega)$ are polynomials with roots $z_{i\ne 0}$ and $p_{i\ne 0}$ respectively. For the near-resonance response of high-Q resonators, $\bar{N}(\omega)$ and $\bar{D}(\omega)$ are approximately constants within range, $\omega_0' - \delta\omega \lesssim \omega \lesssim \omega_0' + \delta\omega$, where $\delta \omega = \omega_0'/Q$. They can be well approximated with their values at $\omega = \omega_0'$. Therefore, the response function is reduced to
\be
\tilde R(\omega) = \frac{\bar{N}(\omega_0')}{\bar{D}(\omega_0')}\frac{(\omega - z_0)}{(\omega - p_0)}. \label{eq:Romega}
\ee
This approximate response function $\tilde R(\omega)$ only depends on the zero and pole $z_0$ and $p_0$ in addition to a complex scaling factor $\frac{\bar N(\omega_0')}{\bar D(\omega_0')}$. The response function $\tilde R(\omega)$ is asymmetric in general while a symmetric response function requires $\Re[z_0] = \Re[p_0]$ (see Fig.~\ref{fig:pz}). We note that any experimental measured frequency response, e.g. S-parameters, of a high-Q resonator can be fitted using Eq.~\eq{Romega} with three complex numbers, the zero, the pole and the complex scaling factor, as the fitting parameters. We also note that the approximation we applied in Section~\ref{sec:S21}, \ref{sec:VVin} and \ref{sec:asymmetry} for reduced expressions for $S_{21}$ and $V/V_{\t{in}}^+$ are equivalent to finding the approximate form of the closest pole $p_0$ and zero $z_0$ in Eq.~\eq{Romega}.

For distributed resonators such as coplanar waveguide resonators, the response function is not simply a ratio of polynomials of $\omega$ but also depends on a wave propagation term $z = e^{j \omega \tau}$, where $\tau$ corresponds to the electrical length of the waveguide. The general response function can be written as
\be
R(\omega) = \frac{N(\omega, z)}{D(\omega, z)},
\ee
where $N(\omega, z)$ and $D(\omega, z)$ are polynomials of both $\omega$ and $z$. Provided $\delta\omega \times \tau \ll 2\pi$, we can take the first order expansion of $z$ as $z \approx e^{j \omega_0' \tau} + j \tau e^{j \omega_0' \tau} (\omega - \omega_0')$. Thus we approximate the near-resonance response of a certain harmonic of the distributed resonator into a ratio of polynomials. Then, the same procedure to reduce the response function to the linear fractional form in Eq.~\eq{Romega} can be applied. As a result, the CPZM is applicable to distributed resonators as well.

\section{Conclusion}
We developed the Closest Pole \& Zero Method (CPZM) for analyzing the near-resonance response of superconducting resonators. We showed that in the high-Q limit, the response functions are well described by a linear fractional form. We first considered the transmission S-parameter ($S_{21}$) of a lumped resonator. Using the CPZM, we obtained a linear fractional $S_{21}$ of this circuit. By fitting the linear fractional $S_{21}$, one can extract the resonance frequency, internal quality factor, and external quality factor out of the experimental measured transmission. We then applied the CPZM to analyze the relation between the capacitor voltage $V$ and the excitation voltage $V_{\t{in}}^+$ of the same circuit. We found that the ratio between $V$ and $V_{\t{in}}^+$ depends on all the fitting parameters of the transmission function in particular the internal quality factor. Since the internal quality factor depends on $V$ itself due to the microscopic mechanism of the loss, $V$ is not proportional to $V_{\t{in}}^+$ in general. We presented a self-consistent method to calculate $V$ from the fitting results to the transmission S-parameter for every given $V_{\t{in}}^+$.

Moreover,we studied the case in which imperfections of the transmission lines used for S-parameter measurements are modeled by introducing impedance mismatch and inductance from wire bonds. We found that the asymmetric transmission near the resonance is mainly caused by wave reflection on the wire bond positions. Using the CPZM, we showed that our fitting routine for extracting the internal quality factor applies to this non-ideal case as well.

We also discussed the general applications of the CPZM to high-Q resonance circuits. We showed that the response function of an arbitrary lumped or distributed resonator is or can be approximated by a ratio of polynomials. The ratio of polynomials can be approximated further to a linear fractional form by finding its pole and zero closest to the bare resonance frequency of the resonator. We described the general procedure for such approximations.

The methods presented here provide a systematic and rigorous treatment of response functions of high quality factor resonators. They allow the reliable determination of the quality factor and electric field in driven resonators. These methods are applicable to a variety of experimental investigations of the mechanisms of dielectric loss in the quantum regime.

\begin{acknowledgments}
We acknowledge Jean-Luc Orgiazzi, Mustafa Bal, and Florian Ong for the help with the experiments. This work is supported by NSERC, CFI, Industry Canada, Ministry of Research and Innovation and  CMC Microsystems. CD is supported by an Ontario Graduate Scholarship. AL is supported by a Sloan Fellowship and an Early Researcher Award.
\end{acknowledgments}

\appendix
\section{} \label{sec:appendix}
This appendix contains the notation definitions and details of approximations of $S_{21}$ and $V/V_{\t{in}}^+$ in Sec.~\ref{sec:S21} and \ref{sec:VVin}.

We first consider the response functions, $S_{21}$ and $V/V_{\t{in}}^+$ (Eq.~\eq{S21} and \eq{VVin}), of the circuit model shown in Fig.~\ref{fig:sample} b) and replace the circuit parameters in these expressions with the reduced parameters using Eq.~\eq{dimlessPara}. After performing parameter replacements and expression simplifications in \emph{Mathematica}\cite{mathematica}, we obtain the following polynomial fractional forms for $S_{21}$ and $v = V/V_{\t{in}}^+$:
\ba
S_{21} =& \frac{a_0 + a_1 q x+ a_2 (q x)^2}{b_0 + b_1 q x+ b_2 (q x)^2 + b_3 (q x)^3 + b_4 (q x)^4}, \\
v =& \frac{c_0 + c_1 qx + c_2 (qx)^2}{d_0 + d_1 qx + d_2 (qx)^2 + d_3 (qx)^3 + + d_4 (qx)^4}. \label{eq:VVinred}
\ea
Their coefficients are defined as
\ba
a_0 =& +32 (-j q+\gamma -\alpha  \gamma ) \xi\\
a_1 =& +32 (1+\gamma -\alpha  \gamma ) \xi\\
a_2 =& +8 (1+\gamma -\alpha  \gamma ) \xi \\
c_0 =& +32 \xi  (\gamma -\alpha  \gamma +j \alpha  \xi ) \\
c_1 =& +16 \xi  (-2 (-1+\alpha ) \gamma +j \alpha  \xi ) \\
c_2 =& -8 (-1+\alpha ) \gamma  \xi
\ea
and
\begin{widetext}
\ba
b_0 = d_0 =& -16 i \left(j q \gamma +\left(i (2+(-2+\alpha ) \alpha ) \gamma +q \left(2+\alpha ^2 \gamma -\beta  \gamma \right)\right) \xi +\left(j \beta  (q+j \gamma )+\alpha ^2 (1-j q+\gamma )\right) \xi ^2\right)\\
b_1 = d_1 =& +8 \left(2 j \gamma +4 \xi +2 (2+2 (-1+\alpha ) \alpha -\beta ) \gamma  \xi -j \left(3 \alpha ^2 (1+\gamma )-\beta  (2+3 \gamma )\right) \xi ^2\right)\\
b_2 = d_2 =& +4 \left(3 j \gamma +\left(2+\left(2-2 \alpha +6 \alpha ^2-5 \beta \right) \gamma \right) \xi -3 j \left(\alpha ^2-\beta \right) (1+\gamma ) \xi ^2\right)\\
b_3 = d_3 =& +2 j \gamma +8 \left(\alpha ^2-\beta \right) \gamma  \xi -2 j \left(\alpha ^2-\beta \right) (1+\gamma ) \xi ^2\\
b_4 = d_4 =& +\left(\alpha ^2-\beta \right) \gamma  \xi.
\ea
\end{widetext}

We then extract the coefficients of the zeroth and first order $qx$ terms from the corresponding numerator and denominator of each response function, which allows us to rewrite the approximate response function in the linear fractional form. We obtain:
\bg
\tilde{S}_{21} = \eta \times \frac{a + j x}{b + j x}, \\
\t{and} ~ \tilde v = \zeta \times \frac{a' + jx}{b' + jx},
\eg
where
\bg
a = \frac{q-j (-1+\alpha ) \gamma }{q (1+\gamma -\alpha  \gamma )}, \\
a' = \frac{2 j (-1+\alpha ) \gamma +2 \alpha  \xi }{2 q (-1+\alpha ) \gamma -j q \alpha  \xi},
\eg
and
\begin{widetext}
\ba
b = b' =& \frac{-2 q \gamma +2 \left(-(2+(-2+\alpha ) \alpha ) \gamma +j q \left(2+\alpha ^2 \gamma -\beta  \gamma \right)\right) \xi +2 \left(-\beta  (q+j \gamma )+\alpha ^2 (q+j (1+\gamma ))\right) \xi ^2}{q \left(-2 \gamma +2 j (2+(2+2 (-1+\alpha ) \alpha -\beta ) \gamma ) \xi +\left(3 \alpha ^2 (1+\gamma )-\beta  (2+3 \gamma )\right) \xi ^2\right)}, \\
\zeta =& \frac{-4 j (-1+\alpha ) \gamma  \xi -2 \alpha  \xi ^2}{-2 \gamma +2 j (2+(2+2 (-1+\alpha ) \alpha -\beta ) \gamma ) \xi +\left(3 \alpha ^2 (1+\gamma )-\beta  (2+3 \gamma )\right) \xi ^2}.
\ea
\end{widetext}

We rewrite the linear fractional $\tilde S_{21}$ and $\tilde v$ by separating $q$ from the other reduced parameters, which gives:
\be
\tilde S_{21} = \eta \frac{A + j \frac{B}{q} + jx}{C + \frac{D}{q} + j (E+\frac{F}{q}) + jx},
\ee
and
\be
\tilde v = \zeta \frac{\frac{K}{q} + j \frac{J}{q} + jx}{C + \frac{D}{q} + j (E+\frac{F}{q}) + jx}.
\ee
By Taylor expanding the coefficients $A$, $C$, and $E$ in terms of small parameters $\alpha$ and $\gamma$, we find that $A \approx 1$, $C \approx 1$, and $E \approx 0$. We identify the resonance frequency as the minimum value of $|\tilde{S}_{21}|$. By taking the first order derivative of $|\tilde{S}_{21}|$ with $A = 1$, $C = 1$, and $E = 0$, we find the extrema:
\begin{widetext}
\be
x_0' = \frac{-B^2+D^2+F^2+2Dq\pm\sqrt{(D^2+(B-F)^2)((B-F)^2+(D+2q)^2)}}{2(B-F)q}.
\ee
Only the extremum with ``$-$'' sign corresponds to the minimum position. As a result, the true resonance frequency of the circuit is
\be
\omega_0 = \omega_0' \left(1+\frac{-B^2+D^2+F^2+2Dq - \sqrt{(D^2+(B-F)^2)((B-F)^2+(D+2q)^2)}}{2(B-F)}\right).
\ee
\end{widetext}
We take the resonance frequency as the resonance frequency without internal loss ($q=0$) as $\omega_0 = \omega_0' (1-B/2)$. Therefore, the response functions $\tilde S_{21}$ and $\tilde v$ can be rewrite in reduced forms in terms of the resonance frequency $\omega_0$ and internal quality factor $Q_i = \frac{1}{q_i} = \omega_0 R C_0 = \frac{1-B/2}{q}$ as
\be
\tilde{S}_{21,\t{red}} = \eta \frac{1+2j \frac{\omega - \omega_0}{\omega_0 q_i}}{1+\frac{D}{q_i(1-B/2)}+j\frac{F-B}{q_i(1-B/2)}+2j \frac{\omega - \omega_0}{\omega_0 q_i}}
\ee
and
\be
\tilde v_{\t{red}} = \zeta\frac{\frac{K}{q_i(1-B/2)}+\frac{j(J-B)}{q_i(1-B/2)}+2j\frac{\omega - \omega_0}{\omega_0 q_i}}{1+\frac{D}{q_i(1-B/2)}+j\frac{F-B}{q_i(1-B/2)}+2j\frac{\omega - \omega_0}{\omega_0 q_i}}.
\ee
In the end, we calculate the quantity of the reduced response function $\tilde v_{\t{red}}$ for $\omega = \omega_0$. We obtain:
\be
\tilde v_0 = \zeta\frac{\frac{K}{q_i(1-B/2)}+\frac{j(J-B)}{q_i(1-B/2)}}{1+\frac{D}{q_i(1-B/2)} +\frac{j(F-B)}{q_i(1-B/2)}}.
\ee
The quantity of $\tilde v_0$ depends on the values of the parameters which appear in the expression of $\tilde S_{21, \t{red}}$ and a linear scaling factor $\lambda$. We define $\lambda$ as
\be
\lambda = \zeta\frac{K+j(J-B)}{1-B/2}.
\ee
The full expression of $\lambda$ in terms of the reduced parameters is given below:
\begin{widetext}
\be
\lambda = \frac{4 j \xi  (2 \alpha  \xi -(-1+\alpha ) \gamma  (-2 j+\alpha  \xi ))}{(-2+(-1+\alpha ) \gamma ) \left(-2 \gamma +2 j (2+(2+2 (-1+\alpha ) \alpha -\beta ) \gamma ) \xi +\left(3 \alpha ^2 (1+\gamma )-\beta  (2+3 \gamma )\right) \xi ^2\right)}. \label{eq:lambda_full}
\ee
\end{widetext}
The calculations involving replacing parameters, simplifying expressions, extracting coefficients from polynomials, and expanding expressions into Taylor series in this appendix are performed in \emph{Mathematica}\cite{mathematica}.

\end{document}